\documentclass[aip,jmp,amsmath,amssymb,reprint]{revtex4-1}
\usepackage{graphicx}
\usepackage{dcolumn}
\usepackage{bm}
\usepackage{chemformula}

\begin{document}

\newcommand{\bq}{\ensuremath{\mathbf{q}}}
\newcommand{\bp}{\ensuremath{\mathbf{p}}}

\title{Quantum kinetic energy and isotope fractionation in aqueous ionic solutions}

\author{Lu Wang}
\affiliation{Department of Chemistry and Chemical Biology, Rutgers University, Piscataway, New Jersey 08854, USA}
\email{lwang@chem.rutgers.edu}
\author{Michele Ceriotti}
\affiliation{Laboratory of Computational Science and Modelling, Institute of Materials, \'Ecole Polytechnique  F\'ed\'erale de Lausanne, 1015 Lausanne, Switzerland}
\author{Thomas E. Markland}
\affiliation{Department of Chemistry, Stanford University, Stanford, California 94305, USA}

\begin{abstract}
At room temperature, the quantum contribution to the kinetic energy of a water molecule exceeds the classical contribution by an order of magnitude. The quantum kinetic energy (QKE) of a water molecule is modulated by its local chemical environment and leads to uneven partitioning of isotopes between different phases in thermal equilibrium, which would not occur if the nuclei behaved classically. In this work, we use {\it ab initio} path integral simulations to show that QKEs of the water molecules and the equilibrium isotope fractionation ratios of the oxygen and hydrogen isotopes are sensitive probes of the hydrogen bonding structures in aqueous ionic solutions. In particular, we demonstrate how the QKE of water molecules in path integral simulations can be decomposed into translational, rotational and vibrational degrees of freedom, and use them to determine the impact of solvation on different molecular motions. By analyzing the QKEs and isotope fractionation ratios, we show how the addition of the Na$^+$, Cl$^-$ and HPO$_4^{2-}$ ions perturbs the competition between quantum effects in liquid water and impacts their local solvation structures.
\end{abstract}

\maketitle

\section{Introduction}
At thermal equilibrium, isotopes of an element can partition differently between two  phases of matter or chemical environments. This phenomenon is known as isotope fractionation and has found uses in many fields, such as in geochemistry to characterize material circulation on the earth's surface,\cite{Hoefs2009,Bradley2014} and in biochemistry to assess hydrogen bond strengths.\cite{Cleland1998,Cleland1994,Lin1998a,Bao1999,Mildvan2002,Oltrogge2015} The isotope fractionation ratio of an element, 10$^3\ln\alpha$, is directly related to the change in quantum kinetic energy (QKE) of the isotopes upon going from one phase to another.\cite{ceri-mark13jcp} Since the kinetic energies of classical particles are independent of their local environment, equilibrium fractionation arises \emph{entirely} from the quantum mechanical nature of the nuclei.

The fractionation of hydrogen (H) and oxygen ($^{16}$O) and their heavier isotopes, deuterium (D) and $^{18}$O, between liquid water and its vapor are of particular interest. These processes occur as part of the evaporation-condensation equilibrium between the ocean and the atmosphere and have been utilized to track the temperature in geological history.\cite{Hoefs2009,Bradley2014} Since there are essentially no interactions between water molecules in the gas phase, the liquid-vapor fractionation ratio probes the changes in their QKEs in the presence of intermolecular interactions, in particular hydrogen bonds, in the condensed phase. For example, the H/D liquid-vapor fractionation ratio has been shown to result from a delicate balance between two competing quantum effects.\cite{habe+09jcp,li+11pnas,mark-bern12pnas,roma+13jpcl,Wang2014,McKenzie2014,Fang2016,Ceriotti2016} While nuclear quantum effects (NQEs) allow the protons to delocalize along the hydrogen bonds in liquid water, thus decreasing the fractionation ratio, the protons become more confined in the orthogonal directions, giving rise to the opposite effect.\cite{mark-bern12pnas,Wang2014,Ceriotti2016} At 300~K, these two effects almost perfectly cancel each other and hence the net influence of NQEs is small on many properties of liquid water.\cite{habe+09jcp,li+11pnas,mark-bern12pnas,roma+13jpcl,Ceriotti2016,Marsalek2017} 

From a series of experiments, it has been well established that adding salts to liquid water alters its hydrogen and oxygen fractionation ratios and the impact strongly depends on the nature of the cations and anions.\cite{Truesdell1974,Stewart1975,Horita2004,Hoefs2009} For example, the $^{16}$O/$^{18}$O fractionation ratio of water  increases in the presence of structure-breaking ions and decreases with the structure-making ones.\cite{Oneil1991} As such, the isotope fractionation ratios of aqueous solutions can be used to probe the solvation environment of ions, which is of fundamental importance in chemistry, geochemistry and biology.\cite{Zhang2006,Hoefs2009,Marcus2009,Brini2017} While classical molecular simulations and {\it ab initio} molecular dynamics (AIMD) simulations are powerful tools to explore the structure and dynamics of aqueous ionic solutions,\cite{Ohtaki1993,Grossfield2003,Fennell2009,Marcus2009,Brini2017,Ramaniah1999,Raugei2001,Lightstone2001,Lyubartsev2001,Liu2010,Hassanali2014} they treat the nuclei as classical particles and hence cannot correctly describe the isotope fractionation processes. Path integral molecular dynamics (PIMD) simulations allow NQEs to be exactly included in the calculation of static equilibrium properties, such as QKEs and fractionation ratios, on a given potential energy surface. \cite{chan-woly81jcp,Parrinello1984,bern-thir86arpc} Recent studies have combined path integral simulations and empirical fixed charge force fields to assess how NQEs affect the hydrogen bond and water exchange dynamics around monatomic alkali and halide ions,\cite{Habershon2014,Wilkins2015} the kinetic energy changes they engender in the water molecules around them, \cite{Wilkins2015,Videla2018} and the effect of these changes on the fractionation ratios and infrared absorption spectra of the aqueous solutions.\cite{Videla2018}

In this work, we perform {\it ab initio} path integral molecular dynamics (AI-PIMD) simulations, which provide a quantum mechanical description of both the electrons and nuclei, of liquid water and aqueous solutions containing the monatomic Na$^+$ and Cl$^-$ ions and the polyatomic HPO$_4^{2-}$ ion. Using these simulations, we demonstrate how the QKE of water molecules obtained from path integral simiulations can be decomposed in terms of their translational, rotational and vibrational degrees of freedom (DOFs). We then use this decomposition to examine the competing quantum effects in liquid water and aqueous ionic solutions, and show that the equilibrium isotope fractionation ratios of the oxygen and hydrogen isotopes are sensitive probes of the local hydrogen bonding environment and ion-water interactions.

\section{Theoretical methods}
In this section, we first summarize the methods used to compute the hydrogen and oxygen fractionation ratios between liquid water and its vapor from PIMD simulations (Sec.~\ref{sec:calc_frac}). In Sec.~\ref{sec:mol_decomp}, we show how the QKE of a water molecule obtained from a path integral simulation can be decomposed into components that correspond to its molecular motions. This analysis allows for a transparent interpretation of the kinetic energy differences observed in the solvated species, as presented in Sec.~\ref{sec:results}, which lead to isotope fractionation in aqueous ionic solutions.  

\subsection{Calculating the liquid-vapor fractionation ratio}
\label{sec:calc_frac}
The H/D fractionation ratio, $10^3\ln\alpha$(D) between the liquid (l) and vapor (v) phases arises from the isotope exchange equilibrium, 
\begin{equation*}
\ch{H2O~(l) + HOD~(v) <=> HOD~(l) + H2O~(v)}.
\end{equation*}
Similarly, we refer to the $^{16}$O/$^{17}$O and $^{16}$O/$^{18}$O fractionation ratios as $10^3 \ln\alpha (^{17}\text{O})$ and $10^3 \ln\alpha (^{18}\text{O})$, respectively, and they correspond to the following equilibria,
\begin{equation*}
\ch{H2$^{16}$O~(l) + H2$^{17}$O~(v) <=> H2$^{17}$O~(l) + H2$^{16}$O~(v)}, 
\end{equation*}
\begin{equation*}
\ch{H2$^{16}$O~(l) + H2$^{18}$O~(v) <=> H2$^{18}$O~(l) + H2$^{16}$O~(v)}.
\end{equation*}
The fractionation ratios, $10^3 \ln\alpha$($j$), are proportional to the free energy difference \cite{Ceriotti2016}
\begin{equation}
\label{eq:freeenergy}
10^3 \ln\alpha(j)=-10^3(\Delta A_j^l-\Delta A_j^v)/k_BT,
\end{equation}
where $k_B$ is the Boltzmann constant and $T$ is the temperature. $\Delta A_j^l$ and $\Delta A_j^v$ are the changes in the free energy upon converting the element $j$ from its lighter isotope (H or $^{16}$O) to the heavier isotope (D, $^{17}$O or $^{18}$O) in the liquid and vapor phases, respectively. In turn, these free energy changes are related to the QKEs of the atoms in each phase,\cite{ceri-mark13jcp,Wang2014} 
\begin{equation}
\Delta A_j^i=-\int_{m_j}^{m_j^\prime} \frac{\langle T_j^i(\mu)\rangle}{\mu}d\mu .
\label{eq:FE}
\end{equation}
Here $m_j$ and $m_j^\prime$ are the masses of the lighter and heavier isotopes of element $j$, respectively. $\langle T_j^i(\mu)\rangle$ is the average QKE of an isotope of mass $\mu$ in the phase $i$. 

The average QKE in Eq.~\ref{eq:FE} can be computed using PIMD simulations, which exactly include NQEs for static equilibrium properties of systems of distinguishable particles by exploiting the isomorphism between a quantum mechanical system and a classical system of ring polymers.\cite{feyn-hibb65book,chan-woly81jcp,Parrinello1984,bern-thir86arpc} If a quantum mechanical system contains $N$ particles with the set of masses $\{ m_j \}$, the ring polymer Hamiltonian in the PIMD simulation is\cite{chan-woly81jcp,bern-thir86arpc}
\begin{eqnarray}
H_P({\bp},{\bq}) = &&\sum_{k=1}^P \left( \sum_{j=1}^N {\frac{|{\bp}_{j}^{(k)}|^2}{2 m_j}} +  \frac{1}{2}m_j\omega_P^2 ({\bq}_{j}^{(k)}-{\bq}_{j}^{(k-1)})^2 \right)\nonumber \\ && +  \sum_{k=1}^P V({\bq}^{k}).
\end{eqnarray}
Here each particle is represented by $P$ ring polymer beads, and cyclic boundary conditions, $k + P \equiv k$, are implied. $\bq_j^{(k)}$ and $\bp_j^{(k)}$ are the position and momentum of the $k^{th}$ bead of particle $j$, respectively. $\omega_P=Pk_BT/\hbar$, and V({$\bq^k$}) is the potential energy of the system. From PIMD simulations, the average kinetic energy of the $j^{th}$ particle can be obtained using the centroid virial estimator,\cite{herm-bern82jcp,cao-bern89jcp}
\begin{equation}
\label{eq:centroid_virial}
\langle T_j\rangle=\left< \frac{3}{2}k_BT+\frac{1}{2P}\sum_{k=1}^{P}({\bq_j^{(k)}}-{\bf \bar{q}_j})\cdot\frac{\partial V ({\bq}^{k})}{\partial \bq_j^{(k)}}\right>,
\end{equation}
where $\bar{\bq}_j=\sum_{k=1}^P \bq_j^{(k)}/P$ is the centroid position of the ring polymer representing particle $j$.  

\subsection{Molecular decomposition of the quantum kinetic energy}
\label{sec:mol_decomp}
The QKE of a molecule can be decomposed into elements that correspond to the translational, rotational and vibrational DOFs. To perform the decomposition, one first constructs the molecular kinetic energy tensor 
\begin{eqnarray}
\label{eq:MolDecomp}
\langle T_{i{\alpha}j\beta}\rangle=&&\left< \frac{k_BT}{2}\delta_{i\alpha,j\beta}+\frac{1}{4P}\sum_{k=1}^{P} \left[\sqrt{\frac{m_i}{m_j}}(q_{i\alpha}^{(k)}-\bar{q}_{i\alpha})\frac{\partial V}{\partial q_{j\beta}^{(k)}}
\right.\right.\nonumber\\
&&\left.\left. 
+\sqrt{\frac{m_j}{m_i}}(q_{j\beta}^{(k)}-\bar{q}_{j\beta})\frac{\partial V}{\partial q_{i\alpha}^{(k)}}\right] \right>.
\end{eqnarray}
Here $i$ and $j$ index the atoms in the molecule, while $\alpha$ and $\beta$ correspond to the x, y and z components of their coordinates. Hence, $q_{i\alpha}^{(k)}$ is the coordinate of the $k^{th}$ ring polymer bead of atom $i$ in the $\alpha$ direction. For example, in the case of a water molecule, $i$ and $j$ represent the O and H atoms and $q_{11}^{(3)}$ is the coordinate of the 3$^{rd}$ ring polymer bead of the O atom in the x direction. When $i=j$ and $\alpha=\beta$, Equation~\ref{eq:MolDecomp} reduces to the centroid virial estimator of a single particle in one direction.

To apply Eq.~\ref{eq:MolDecomp}, one must consider that molecules are able to rotate in the liquid. Therefore, the average kinetic energy matrix $\langle T_{i{\alpha}j\beta}\rangle$ is only physically meaningful if each molecule is aligned to a common reference frame. The molecular decomposition of the QKE can then be achieved by diagonalizing the resulting kinetic energy matrix, and the eigenvectors give the direction of the molecular motion and the eigenvalues are the corresponding QKE components.

For example, when describing water in the liquid and vapor phases, we used a reference frame that positioned the water molecule on the x--y plane, the O atom on the x-axis and the geometric center of the molecules on the origin, as depicted in the inset of Fig. \ref{fig:moldecomp}. We aligned each water molecule with the reference molecule using the Kabsch algorithm\cite{Kabsch1976} and calculated its $T_{i{\alpha}j\beta}$ from Eq.~\ref{eq:MolDecomp}. The average kinetic energy tensor $\langle T_{i{\alpha}j\beta}\rangle$, a 9$\times$9 matrix, was obtained by averaging over all water molecules in all snapshots from the AI-PIMD simulations. We then diagonalized this matrix to obtain the QKE values corresponding to the translational, rotational and vibrational DOFs of the water molecules in the two phases. 

\section{Simulation methods}

AIMD and AI-PIMD simulations were performed for liquid water, water vapor and aqueous solutions containing Na$^+$, Cl$^-$ or HPO$_4^{2-}$. The simulations were carried out in the canonical ensemble at 300 K using a time step of 0.5~fs. The total simulation lengths were 50~ps for liquid water and the aqueous ionic solutions, and 250~ps for gaseous water. We used the i-PI program\cite{ceri+14cpc,Kapil2019} for the path integral evolution, and the QuickStep module in the CP2K package\cite{vand-krac05cpc} to generate the electronic potential energy surface. Each atom was represented by 6 ring polymer beads using the path integral generalized Langevin equation method.\cite{ceri-mano12prl} The electronic structure of the systems was evaluated using the BLYP exchange correlation functional\cite{beck88pra,lee+88prb} and the Goedecker-Teter-Hutter pseudopotentials.\cite{goed+96prb} The double-zeta split-valence basis set was used with a cutoff of 300 Ry to represent the charge density. For the gas-phase simulation, a water molecule was placed in a cubic box of length 10 \AA, and the Martyna-Tuckerman Poisson solver was applied.\cite{Martyna1999} Liquid simulations were performed with periodic boundary conditions. The simulations of liquid water contained 64 water molecules in a cubic box of length 12.42~\AA. The aqueous solutions contained 1 ion (Na$^+$, Cl$^-$ or HPO$_4^{2-}$) and 128 water molecules in a cubic box with a length of 15.65~\AA.

From the AI-PIMD simulations we calculated the oxygen and hydrogen fractionation ratios using the thermodynamic free energy perturbation (TD-FEP) path integral estimator.\cite{ceri-mark13jcp} This allowed us to obtain $\Delta A_D^l$, $\Delta A_{^{17}O}^l$, $\Delta A_{^{18}O}^l$, $\Delta A_{^{17}O}^v$ and $\Delta A_{^{18}O}^v$ in Equation \ref{eq:FE} from a single PIMD simulation of the most abundant isotopes. To compute $\Delta A_D^v$ with the required accuracy, we performed separate simulations of H$_2$O and HOD in the gas phase and integrated Equation \ref{eq:FE} by using a quasi-harmonic approximation that assumes $\langle T_D^v(\mu)\rangle\propto 1/\sqrt{\mu}$. To decompose the fractionation ratios according to the hydrogen bond environment, we defined that O--H...O$^\prime$ was hydrogen bonded if the oxygen-oyxgen distance $d_{OO^\prime} < 3.5$ \AA~and the angle $\theta_{HOO^\prime} < 30^\text{o}$.\cite{laag-hyne06science} The radial distribution functions (RDFs) in Fig. \ref{fig:RDF} were obtained from the centroid of the ring polymer beads representing the relevant atoms. Accordingly, we determined the hydration layers for Table \ref{tab:fractionation_solvation} and Fig. \ref{fig:IonDecomp} using the distance between the centroids of the ions and water. To validate the simulations, we also computed the RDFs from the AIMD and AI-PIMD simulations (by averaging over the beads of the ring polymers which are the physical observable). The first peak in the Na-O RDFs of the Na$^+$ solution occurred at a distance of 2.42 and 2.38 \AA~from the AI-PIMD and AIMD simulations, respectively, which are in good agreement with the experimental value of 2.38 \AA.\cite{Galib2017} The first peak of the RDFs of the Cl$^-$ solution appeared at a Cl-H distance of 2.13 and 2.18 \AA~from AI-PIMD and AIMD simulations, respectively, consistent with the experimental value of 2.22 \AA.\cite{Bruni2012}

As an alternative to using PIMD simulations, one can approximate the fractionation ratios from AIMD simulations, which treat the nuclei classically, using the $\hbar^2$-expansion method. This method includes the quantum corrections to the classical partition function and energy up to  order $\hbar^2$.\cite{Kirkwood1933,Powles1979,McQuarrie2000,Chialvo2009} In the $\hbar^2$-expansion method, the difference between the quantum and classical free energies of an atom $j$ is\cite{Powles1979}  
\begin{equation}
A-A_{C}= \left\langle\frac{\hbar^2\beta^2}{24m_j} \left(\frac{\partial V}{\partial {\bq_j}}\right)^2\right\rangle+O(\hbar^4).
\end{equation}
Using this equation, one can obtain the $\hbar^2$-expansion approximations to the H/D fractionation ratio as
\begin{equation}
\resizebox{0.49\textwidth}{!}{
	$10^3 \ln\alpha (D) \approx 10^3\frac{\hbar^2\beta^3}{24} \left(\frac{1}{m_H}-\frac{1}{m_D}\right) 
	\left(\left \langle \left(\frac{\partial V^l}{\partial {\bq_H}}\right)^2\right\rangle - \left\langle \left(\frac{\partial V^v}{\partial {\bq_H}}\right)^2\right\rangle\right).
	$} 
\label{eq:hbar}
\end{equation}
Here $-\frac{\partial V^l}{\partial {\bq_H}}$ and $-\frac{\partial V^v}{\partial {\bq_H}}$ can be recognized as the forces on the H atom in the liquid and vapor phases of H$_2$O, respectively. One should note that when the nuclei are treated classically, as in AIMD simulations, the average force experienced by a particle in a simulation of a given phase is independent of its mass and hence,
\begin{equation}
\left\langle\left(\frac{\partial V}{\partial {\bq_H}}\right)^2\right\rangle = \left\langle\left(\frac{\partial V}{\partial {\bq_D}}\right)^2\right\rangle. 
\end{equation}

\section{Results and discussions}
\label{sec:results}

In the following, we first consider NQEs in pure water and utilize the molecular decomposition method introduced in Sec.~\ref{sec:mol_decomp} to demonstrate the competition between quantum effects. We then show how the hydrogen bonding configurations in liquid water can be analyzed using the oxygen and hydrogen fractionation ratios. Next, we elucidate the influence of ions on these quantum mechanical quantities in aqueous ionic solutions.

\subsection{Quantum kinetic energy and isotope fractionation in liquid water}
\begin{figure}[h]
	\centering
	\includegraphics[height=5cm]{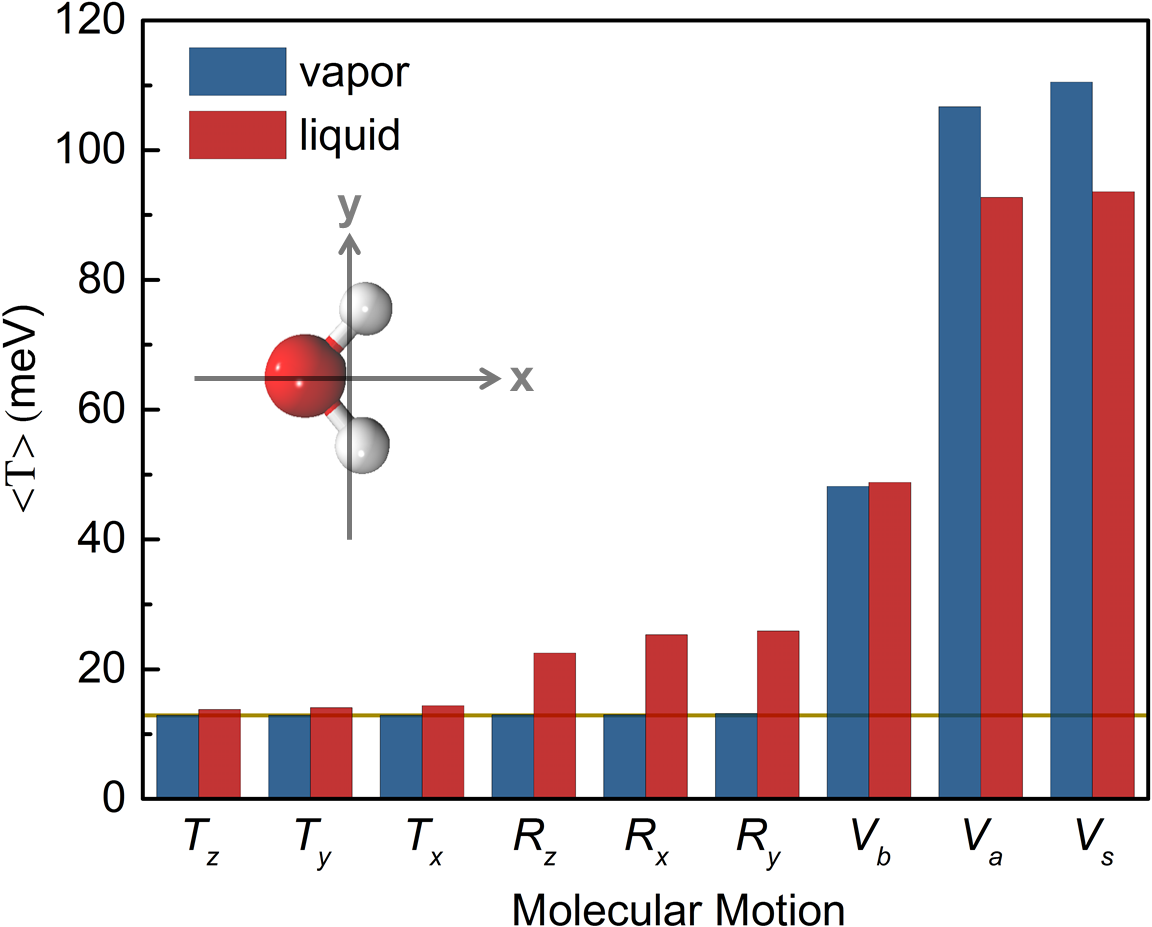}
	\caption{Decomposition of the average QKE of the water molecules in the liquid and gas phases. $T$, $R$ and $V$ stand for translation, rotation and vibration of a water molecule, respectively. $x$, $y$ and $z$ are the directions of molecular motions, and the coordinate system is shown in the inset with $z$ pointing out of the paper plane. V$_b$, V$_a$ and V$_s$ correspond to bending and asymmetric and symmetric stretching DOFs, respectively. The horizontal line represents the classical kinetic energy of $\frac{1}{2}k_BT$, which is 12.9 meV at 300 K.}
	\label{fig:moldecomp}
\end{figure}

Fig.~\ref{fig:moldecomp} shows the results obtained from applying the QKE decomposition method to water molecules in the gas and liquid phases. If the nuclei were classical, the equipartition theorem dictates that each DOF of a water molecule would contribute $\frac{1}{2}k_{B}T$ to the total kinetic energy, which is equal to 12.9 meV at 300 K. As shown in Fig.~\ref{fig:moldecomp}, QKEs in the translational, rotational and vibrational DOFs can differ significantly from the classical predictions.

In the gas phase, the translational QKEs of water are identical to the classical value and those associated with the molecular rotations deviate by less than 0.01\%. However, the average QKEs in the vibrational modes are 3.7 to 8.6-fold larger than $\frac{1}{2}k_{B}T$, demonstrating the quantum mechanical nature of these high-frequency bending (V$_b$) and anti-symmetric (V$_a$) and symmetric (V$_s$) stretching modes.

Unlike the classical kinetic energy, QKEs of a particle are sensitive to its chemical environment, as they increase when the particle is confined along a particular DOF. In liquid water, the formation of hydrogen bonds allows the protons to be more delocalized along the O--H stretching DOFs. Accordingly, the QKEs in $V_a$ and $V_s$ are both reduced by over 14 meV as compared to the corresponding values in the gas phase. This leads to a total reduction of 30.9 meV (1.2$k_{B}T$), as shown in Fig.~\ref{fig:moldecomp}. However, compared with gaseous water, the hydrogen bonding interactions and tight packing in the liquid also hinder the free rotation, bending and translation of the molecules and increase the QKEs in these DOFs by a total amount of 38.8 meV (1.5$k_{B}T$). From these two competing effects, we observe a net increase of 7.9 meV (0.3$k_{B}T$) in the QKE upon moving from gaseous water to liquid water. This 80\% cancellation in QKEs between different DOFs also demonstrates the principle of competing quantum effects, which has been extensively studied in hydrogen bonded systems.\cite{habe+09jcp,li+11pnas,mark-bern12pnas,roma+13jpcl,McKenzie2014,Wang2014,Wang2014a,Ceriotti2016,Fang2016,Cheng2019,Fang2019a}  

Equilibrium isotope fractionation ratios effectively report the differences in the QKE between the isotopes of an element.\cite{Hoefs2009,mark-bern12pnas,ceri-mark13jcp,Pinilla2014,Ceriotti2016,Cheng2016} Considering that the relative abundance of $^{16}$O and its heavier isotopes $^{17}$O and $^{18}$O are important tracers of the Earth hydrological cycle,\cite{Angert2004,Barkan2005} we calculated the liquid-vapor fractionation ratios $10^3\ln\alpha (^{17}\text{O})$ and $10^3\ln\alpha (^{18}\text{O})$. As shown in Table \ref{tab:Ofractionation}, $10^3\ln\alpha (^{17}\text{O})$ and $10^3\ln\alpha (^{18}\text{O})$ obtained from both simulations and experiments\cite{Barkan2005} are positive, indicating that the heavier isotopes $^{17}$O and $^{18}$O are preferentially found in the liquid phase, whereas $^{16}$O favors the vapor phase. This is consistent with the experimental observation that the lighter isotopes prefer to reside in the gas phase in the evaporation and precipitation process under equilibrium conditions.\cite{Hoefs2009} In addition, we computed the ratio $10^3\ln\alpha (^{17}\text{O})$/$10^3\ln\alpha (^{18}\text{O})$, which is widely used to evaluate the triple-isotope systems,\cite{Mook2000,Angert2004,Barkan2005} and found it to be in quantitative agreement with the experimental value.\cite{Angert2004,Barkan2005}

\begin{table}
	\caption{\label{tab:Ofractionation} $^{16}$O/$^{17}$O and $^{16}$O/$^{18}$O fractionation ratios calculated from AI-PIMD simulations. The experimental values are also listed.\cite{Barkan2005} The error bars of the calculated fractionation values are $\pm0.10$.}
	\begin{tabular*}{0.48\textwidth}{@{\extracolsep{\fill}}lcr}
		\hline
		& AI-PIMD & Experiment \\
		\hline
		$10^3\ln\alpha (^{17}$O) & 6.38 & 4.95 $\pm$ 0.02\\
		$10^3\ln\alpha (^{18}$O) & 12.07 & 9.36 $\pm$ 0.02\\
		$10^3\ln\alpha (^{17}\text{O})$/$10^3\ln\alpha (^{18}\text{O})$ & 0.529 & 0.529 $\pm$ 0.001\\
		\hline
	\end{tabular*}
\end{table}

\begin{figure}[h]
	\centering
	\includegraphics[height=1.29cm]{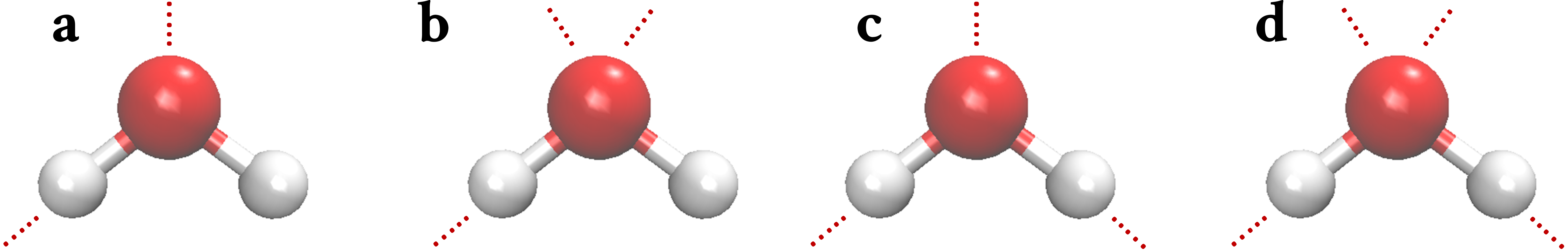}
	\caption{Most abundant hydrogen bonding configurations of a water molecule from our AI-PIMD simulations. Red and white represent the O and H atom, respectively, and the dotted lines are their hydrogen bonds with surrounding water molecules. The configurations contain (a) 1 hydrogen bond donor and 1 acceptor, (b) 1 hydrogen bond donor and 2 acceptors, (c) 2 hydrogen bond donors and 1 acceptor and (d) 2 hydrogen bond donors and 2 acceptors.}
	\label{fig:hbond}
\end{figure} 

From Table \ref{tab:Ofractionation}, our AI-PIMD simulations using the BLYP functional overestimate the values of $10^3\ln\alpha (^{17}$O) and $10^3\ln\alpha (^{18}$O) as compared to those measured in experiments.\cite{Barkan2005} To elucidate the origin of the overestimation, we identified the four main hydrogen bonding configurations from the AI-PIMD simulations of liquid water (Fig.~\ref{fig:hbond}), and decomposed the $10^3\ln\alpha (^{18}$O) value of water based on its hydrogen bonding environment. When an O atom accepts a hydrogen bond from a nearby water molecule, it becomes more confined and the zero-point energy associated with the lighter isotope $^{16}$O increases more prominently than that of the heavier isotope $^{18}$O, making $^{16}$O more energetically favorable to reside in the gas phase. Accordingly, increasing the number of hydrogen bond acceptors in a water molecule is accompanied by an increase in the fractionation ratio, as shown in Table \ref{tab:HbondDecomp}. 
Similarly, the O atom in a water molecule is confined in the direction of the hydrogen bond when the O--H group serves as a donor, and hence water molecules with larger number of hydrogen bond donors have increased $10^3\ln\alpha (^{18}$O) values. As a result, Table \ref{tab:HbondDecomp} suggest that the predicted  $10^3\ln\alpha(^{18}$O) is too high because the BLYP functional tends to overstructure liquid water by forming too many tetrahedral hydrogen bonds,\cite{morr-car08prl,Ceriotti2016,Gasparotto2016a} and the same conclusion holds for the $^{16}$O/$^{17}$O fractionation process. From our previous studies of liquid water, this structuring of the hydrogen bond network and overestimation of the fractionation ratios can be partially alleviated by incorporating exact exchange and dispersion corrections.\cite{Wang2014} Therefore, the liquid-vapor fractionation ratios of the oxygen isotopes are sensitive probes of the hydrogen bond environment in liquid water, and can be used to assess the performance of a density functional to correctly describe the hydrogen bonds.

\begin{table}[h!]
	\caption{\label{tab:HbondDecomp} Probabilities of observing the hydrogen bonding configurations from AI-PIMD simulations of liquid water and the corresponding decomposition of $10^3\ln\alpha (^{18}$O).}
	\begin{tabular*}{0.48\textwidth}{@{\extracolsep{\fill}}cccc}
		\hline
		Donor&Acceptor&Probability&10$^3\ln\alpha (^{18}$O)\\
		\hline
		1 & 1 & 1.0\% & 8.9 \\
		1 & 2 & 4.2\% & 10.6 \\
		2 & 1 & 7.4\% & 11.3 \\
		2 & 2 & 87.2\% & 12.3 \\
		\hline
	\end{tabular*}
\end{table}

We now consider whether the free energy changes in the liquid-vapor isotope fractionation equilibrium can be correctly captured by applying a quantum correction to AIMD simulations, which treat the nuclei classically.\cite{Kirkwood1933,Powles1979,McQuarrie2000} The $\hbar^2$-expansion method has previously been used to calculate the isotope fractionation of hydrogen and oxygen isotopes in liquid water from classical molecular dynamics simulations.\cite{Chialvo2009} By applying the $\hbar^2$-expansion to our AIMD simulations of water in the liquid and gas phases, we find $10^3\ln\alpha$(D) and $10^3\ln\alpha(^{18}$O) are -983 and -14, respectively, at 300 K. They disagree quantitatively and qualitatively with the experimental values (73 and 9.36, respectively\cite{Horita1994,Barkan2005}) and the  AI-PIMD results (62 and 12.07, respectively). In particular, this  approximation leads to the incorrect prediction that lighter isotopes H and $^{16}$O are preferred in the liquid, while the heavier isotopes D and $^{18}$O are preferred in the gas phase. 

To understand the errors in the $\hbar^2$-expansion approximation, we decompose the $10^3\ln\alpha$(D) value predicted by the $\hbar^2$-expansion method into contributions from three orthogonal directions: one along the O--H bond, one in the plane of the water molecule and orthogonal to the O--H direction, and one perpendicular to the molecular plane. As shown in Table \ref{tab:hbar_expansion}, when compared to the predictions from AI-PIMD simulations, which exactly include NQEs, the $\hbar^2$-expansion method overestimates the O--H stretch contribution by almost 5 fold. This is mainly due to the fact that the stretch DOF has large vibrational frequencies and hence is highly quantum mechanical in nature, which is beyond the region of applicability of the expansion. In contrast, the two orthogonal modes have lower frequencies and can be well modeled using the $\hbar^2$-expansion approximation. Because of the overestimation of the O--H contribution, the $\hbar^2$-expansion method does not correctly capture the balance of the competing quantum effects, leading to spuriously inverted fractionation. 

\begin{table}[h!]
	\caption{\label{tab:hbar_expansion} $10^3\ln\alpha$(D) calculated from AIMD simulations and the $\hbar^2$-expansion method, and from AI-PIMD simulations using the TD-FEP method.\cite{Wang2014} The total $10^3\ln\alpha$(D) values are decomposed into contributions from three orthogonal directions: along the O--H bond direction (O--H), in the plane of the water molecule (in plane) and perpendicular to the molecular plane (out of plane).}
	\begin{tabular*}{0.48\textwidth}{@{\extracolsep{\fill}}lcc}
		\hline
		& $\hbar^2$-expansion & AI-PIMD\\
		\hline
		O--H & -1393 & -292 \\
		In plane & 119 & 104\\
		Out of plane & 291 & 250\\
		Total & -983 & 62 \\
		\hline
	\end{tabular*}
	\label{tab:hbar}
\end{table}

\subsection{Nuclear quantum effects in aqueous ionic solutions}
The presence of salts can facilitate or disrupt the hydrogen bond networks in liquid water. To examine how ions alter the liquid-vapor isotope fractionation equilibrium of water, we performed AI-PIMD simulations of aqueous solutions that contain 0.43 M of Na$^+$, Cl$^-$ or HPO$_4^{2-}$, and found the H/D fractionation ratios to be 61, 57 and 50, respectively. Compared to a $10^3\ln\alpha$(D) value of 62 for pure water,\cite{Wang2014} in all 3 cases, the addition of ions reduces the fractionation ratio and makes the heavier isotope D less likely to reside in the liquid phase.

\begin{figure}[h]
	\centering
	\includegraphics[height=4.2cm]{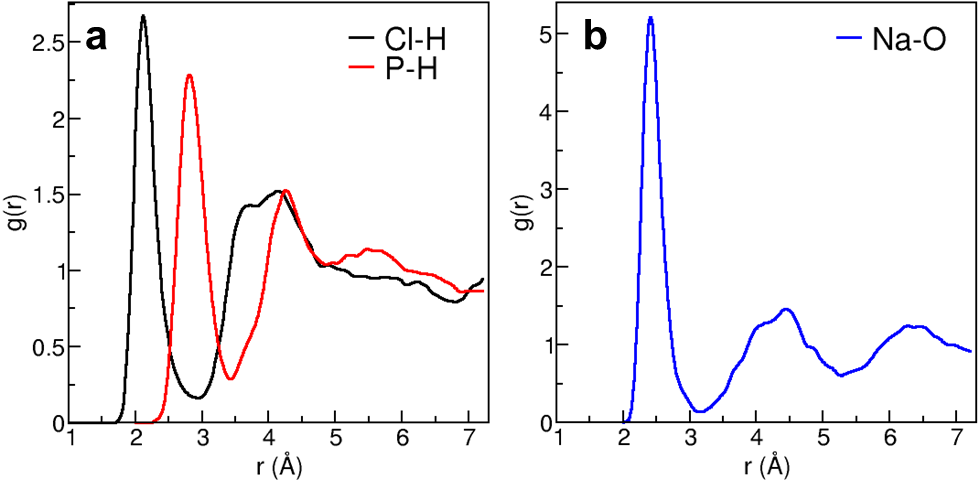}
	\caption{Radial distribution functions between (a) Cl$^-$ and HPO$_4^{2-}$ with the water H's, and (b) Na$^+$ and the water O's from AI-PIMD simulations.}
	\label{fig:RDF}
\end{figure}

Since it has been proposed that the isotope salt effects arise mainly from the different hydration conditions around the ions,\cite{Taube1954,Oneil1991,Horita1994} we calculated the RDFs between the ions and water (Fig. \ref{fig:RDF}) and examined the QKEs of the hydrogen atoms in the first and second hydration shells of the cation and anions. Since the anions Cl$^-$ and HPO$_4^{2-}$ receive hydrogen bonds from the H atoms in water, we defined their first (second) coordination shell using the first (second) minimum in the RDF between the anion atoms (Cl or P) and the solvent H atoms. As shown in Fig. \ref{fig:RDF}a, this gives the ion-H distances of 2.94 and 5.33 \AA~for Cl$^-$, and 3.42 and 4.86 \AA~for HPO$_4^{2-}$ for their first and second hydration layers, respectively. For Na$^+$, we consider the first- and second-shell water molecules as those within the first (3.15 \AA) and second minima (5.28 \AA) of the Na--O RDF, as demonstrated in Fig. \ref{fig:RDF}b. Accordingly, its first- and second-shell hydrogens are the H atoms that belong to these molecules. From the AI-PIMD simulations, we find an average of 10, 5 and 8 first shell hydrogens around Na$^+$, Cl$^-$ and HPO$_4^{2-}$, respectively. In addition, there are an average of 30, 40 and 22 second shell hydrogens for the 3 ions, respectively.

To evaluate the effects of the ions, we computed the differences of the average QKE between the H atoms in the first and second shells of an ion and those in pure water (pw), $\Delta T_{1}=\langle T\rangle_{1}-\langle T\rangle_{pw}$ and $\Delta T_{2}=\langle T\rangle_{2}-\langle T\rangle_{pw}$. As shown in Table \ref{tab:fractionation_solvation}, the presence of an ion significantly alters the QKEs of the solvent molecules in its close proximity. Na$^+$ leads to a positive $\Delta T_1$ of 0.7 meV, whereas Cl$^-$ and HPO$_4^{2-}$ decrease the average QKEs of the first shell hydrogens with $\Delta T_1$ of -2.5 meV and -3.1 meV, respectively. Therefore, the impact of an ion on a nearby H atom's QKE is comparable to the effect of going from the liquid to the vapor phase, which reduces the QKE of the hydrogens by 2.7 meV. As the ion-water distance increases, the behavior of the H atoms approach those in pure water and the magnitude of $\Delta T_2$ in all 3 cases are below 0.7 meV.  

Next, we examine how the ions change the H/D fractionation ratio between liquid and gaseous water. For this purpose, we write $10^3\ln\alpha(D)$ in terms of the average QKE of the H atoms in the liquid and vapor phases, $\langle T_H^l\rangle$ and $\langle T_H^v\rangle$,  using a quasi-harmonic approximation. Assuming that $\langle T_D^l(\mu)\rangle$ and $\langle T_D^v(\mu)\rangle$ both scale as $1/\sqrt{\mu}$,\cite{ceri-mark13jcp} we can simplify Eq.~\ref{eq:FE} to obtain\cite{Cheng2016}
\begin{equation}
\label{eq:quasiharmonic}
10^3\ln\alpha(D)=\frac{2000 \left( 1-\sqrt{\frac{m_H}{m_D}} \right) (\langle T_H^l\rangle-\langle T_H^v\rangle)}{k_BT}. 
\end{equation}
Here $m_H$ and $m_D$ are the masses of H and D atoms, respectively. To validate this approximation, we apply Eq. \ref{eq:quasiharmonic} to pure water and aqueous ionic solutions containing Na$^+$, Cl$^-$ and HPO$_4^{2-}$ and obtain their $10^3\ln\alpha(D)$ of 61, 59, 55 and 49, respectively. They are in good agreement with the values of 62, 61, 57 and 50, respectively, as calculated using the TD-FEP method.

\begin{table}[h!]
	\caption{\label{tab:fractionation_solvation} Changes in QKEs and H/D fractionation ratios of the first- and second-shell hydrogens around the ions as compared to those in pure water. In pure water, $\langle T\rangle_{pw}$ is 148.3 meV\cite{Wang2014} and $10^3\ln\alpha_{pw}$ calculated from the quasi-harmonic approximation is 61.}
	\begin{tabular*}{0.48\textwidth}{@{\extracolsep{\fill}}lcccc}
		\hline
		Ion&$\Delta T_{1}$ (meV) &  $\Delta T_{2}$ (meV)& $\Delta10^3\ln\alpha_{1}$ & $\Delta10^3\ln\alpha_{2}$\\
		\hline
		Na$^+$ & +0.7 & -0.4 & +17  & -8\\
		Cl$^-$ & -2.5 & -0.4 & -56  & -10 \\
		HPO$_4^{2-}$ & -3.1 & -0.7 & -69 & -15 \\
		\hline
	\end{tabular*}
\end{table}

From Eq.~\ref{eq:quasiharmonic}, we use $\Delta T_{1}$ and $\Delta T_2$ to compute fractionation ratios of the H atoms in the ions' hydration layers relative to those in pure water, $\Delta 10^3\ln\alpha_{1}=10^3\ln\alpha_{1}(D)-10^3\ln\alpha_{pw}(D)$ and $\Delta 10^3\ln\alpha_{2}=10^3\ln\alpha_{2}(D)-10^3\ln\alpha_{pw}(D)$. These $\Delta 10^3\ln\alpha$ describe the isotope exchange equilibrium between the hydration layers of the ions and bulk water in the aqueous ionic solutions,
\begin{equation*}
\ch{H2O~(hydr) + HOD(bulk) <=> HOD~(hydr) + H2O~(bulk)}
\end{equation*}
As shown in Table \ref{tab:fractionation_solvation}, hydrogens in the first coordination shell of Na$^+$ have increased QKEs ($\Delta T_1=$+0.7 meV) and accordingly their $\Delta 10^3\ln\alpha_1$ is +17. This suggests that the heavier D atoms are more likely to reside in the vicinity of the cation, whereas the H atoms prefer to be in the bulk of the solution. In contrast, both Cl$^-$ and HPO$_4^{2-}$ reduce the QKEs of the first shell hydrogens and result in negative $\Delta 10^3\ln\alpha_1$ values of -56 and -69, respectively. Going beyond the first coordination shell, $\Delta 10^3\ln\alpha_2$ of all the cations and anions are negative, indicating that H, rather than D, is favored in the second hydration layers of these ions.

\begin{figure}[h]
	\centering
	\includegraphics[height=5.3cm]{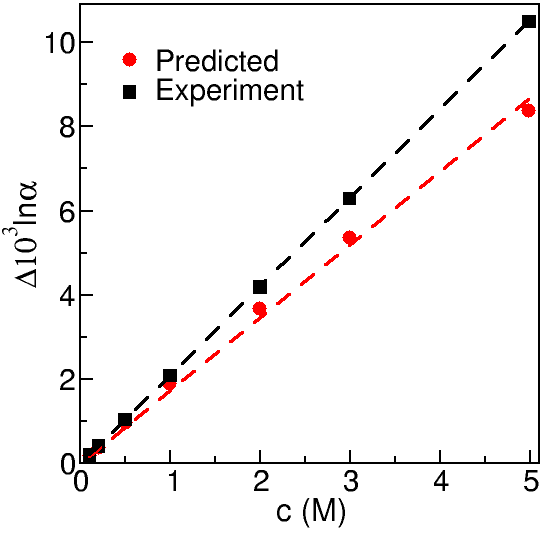}
	\caption{$\Delta 10^3\ln\alpha$ predicted from Eq. \ref{eq:frac} and from experiment\cite{Horita2004} for different concentrations of the NaCl solution at 300 K.}
	\label{fig:Linear}
\end{figure}

Since the ions exert the strongest impact on the first layer hydrogen atoms, one can combine the $\Delta 10^3\ln\alpha_1$ values of a cation and an anion and their average coordination numbers to calculate the H/D fractionation ratio of a salt solution relative to that of pure water, 
\begin{equation}
\label{eq:frac}
\begin{split}
\Delta 10^3\ln\alpha&=10^3\ln\alpha(D)_{soln}-10^3\ln\alpha(D)_{pw}\\
&=x_1^{cation}[\Delta 10^3\ln\alpha_1^{cation}]+x_1^{anion}[\Delta 10^3\ln\alpha_1^{anion}].
\end{split}
\end{equation}
Here $x_1^{cation}$ and $x_1^{anion}$ are the mole fractions of the first-shell hydrogens around the cation and anion, respectively. In pure water, $x_1^{cation}=x_1^{anion}=0$ and hence $\Delta 10^3\ln\alpha$ is 0. As an example of applying Eq. \ref{eq:frac}, we consider a 0.2 M NaCl solution. As there are 10 and 5 first shell hydrogens around Na$^+$ and Cl$^-$, respectively, $x_1^{cation}$ is 0.036 and $x_1^{anion}$ is 0.018. From Eq. \ref{eq:frac}, $\Delta10^3\ln\alpha$ is -0.40, in good agreement with the experimental value of -0.42 at 300 K.\cite{Horita2004} 
Using Eq. \ref{eq:frac}, we compute $\Delta 10^3\ln\alpha$ for the NaCl solution and find that it follows an almost linear relation with the salt concentration (in M) with a slope of 1.73. As demonstrated in Fig. \ref{fig:Linear}, the predictions from Eq. \ref{eq:frac} are in good agreement with the experimental measurements, which have a slope of 2.11 with respect to the salt concentration.\cite{Horita2004} As such, Eq. \ref{eq:frac} provides a simple and physically transparent way to predict how salts change the H/D fractionation ratio in aqueous solutions based purely on the first solvation shell information, which can effectively explain the experimental observations and guide the design of new experiments to examine the salt effects on the equilibrium isotope distributions between the liquid and vapor phases.

\begin{figure}[h]
	\centering
	\includegraphics[height=13cm]{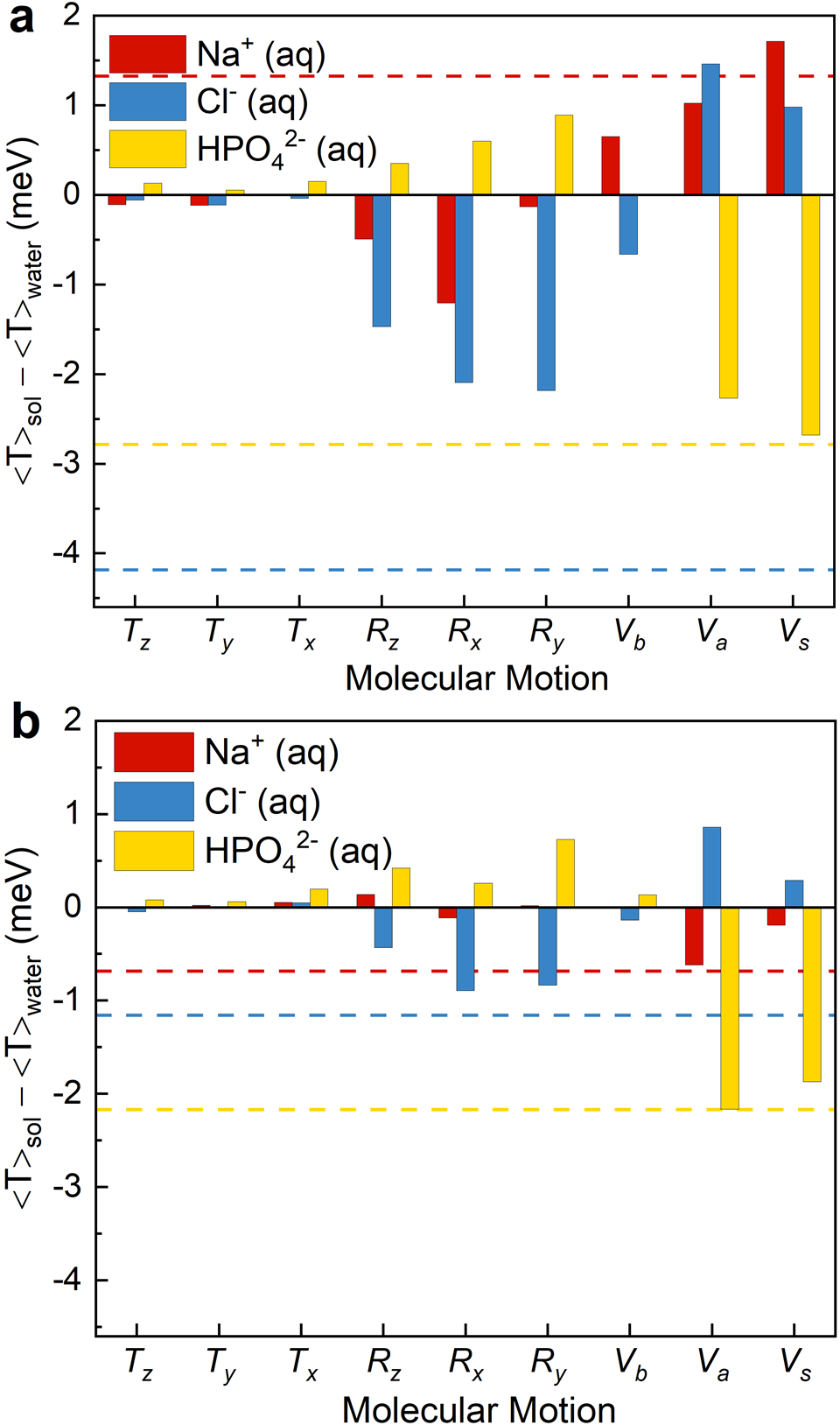}
	\caption{Decomposition of the average QKEs of water molecules in the (a) first and (b) second hydration layers of Na$^+$, Cl$^-$ and HPO$_4^{2-}$, with the values for pure water subtracted from each component. The horizontal lines are the total change in QKE in the (a) first and (b) second solvation shells for the three ions.}
	\label{fig:IonDecomp}
\end{figure}

To further analyze how the ions impact the QKE of the surrounding water molecules, we defined a water molecule to be in an ion's first (second) hydration layers if at least one of the H atoms belong to its first (second) coordination shell. We then decomposed the average QKEs of these hydration water into the translational, rotational and vibrational DOFs using the procedure introduced in Sec.~\ref{sec:mol_decomp}, and present the results in Fig.~\ref{fig:IonDecomp} after subtracting the corresponding QKE values for pure liquid water. In the first hydration layer of Na$^+$, the water molecules are aligned with their O atoms facing the cation. The hydrogen bond network of these water molecules is disrupted, with an average of 2.9 hydrogen bonds per molecule as compared to that of 3.9 in pure water. As shown in Fig.~\ref{fig:IonDecomp}a, this perturbation to the hydrogen bonds makes the O--H bonds in water more confined, enhancing the QKEs associated with their stretching and bending modes. Correspondingly, it also allows the water molecules to rotate and translate more freely, reducing the corresponding QKE elements as compared to those in pure water. These two competing quantum effects largely cancel each other, with the overall QKE of the first layer water molecule increasing by 1.3 meV around the Na$^+$ ion. This observation is consistent with the positive values of $\Delta T_1$ and $\Delta 10^3\ln\alpha_1$ for the first-shell H atoms in Table \ref{tab:fractionation_solvation}. From Fig.~\ref{fig:IonDecomp}b, Na$^+$ has a much smaller influence on its second hydration layer, although it slightly enhances the hydrogen bonding structure of water and gives an average of 3.94 hydrogen bonds per molecule. As a result, the overall QKE of the second layer water molecules decreases by 0.7 meV, and both $\Delta T_2$ and $\Delta 10^3\ln\alpha_2$ become negative (Table \ref{tab:fractionation_solvation}).

The Cl$^-$ ion forms weaker hydrogen bonds with water than those between water molecules. As demonstrated in Fig. \ref{fig:IonDecomp}a, this makes the water molecules more confined in the stretch and bending DOFs and increases their QKE elements. It also facilitates the rotation and translation of the solvent and reduces their corresponding QKEs. The net effect of Cl$^-$ is a reduction of the average QKE of the first layer water by 4.2 meV as compared to pure water. Comparing Figs. \ref{fig:IonDecomp}a and b, the influence of Cl$^-$ on the second hydration layer mimics that for the first layer, although the overall change in the average QKE decreases by 71\%.  Accordingly, $\Delta T_2$ and $\Delta 10^3\ln\alpha_2$ have much smaller magnitude than $\Delta T_1$ and $\Delta 10^3\ln\alpha_1$, respectively, as shown in Table \ref{tab:fractionation_solvation}. 

Compared to the monatomic ions, HPO$_4^{2-}$ possesses a higher charge and is capable of forming directional hydrogen bonds with the solvent. As such, its hydration layer exhibits different behavior compared to the other ions, as shown in Figs. \ref{fig:IonDecomp}a and b. The O--H stretches in the first layer water facilitate the hydrogen bonding interactions between the solvent and the O atoms in HPO$_4^{2-}$, referred to as O$_{HP}$, which allow the proton to be quantum mechanically delocalized along the hydrogen bond and become less confined. From Fig. \ref{fig:IonDecomp}a, QKEs in the stretch DOFs of these water molecules are significantly reduced as compared to pure water. The rotational DOFs counteract this effect by weakening the water-ion hydrogen bonds and increasing the corresponding QKE elements. Cancellation of the competing quantum effects leads to an overall decrease of the QKE of the water molecules by 2.8 meV as compared to those in pure water. Around the large anion HPO$_4^{2-}$, NQEs act to strengthen the O--H$\cdot\cdot\cdot$O$_{HP}$ hydrogen bond to such a degree that the $10^3\ln\alpha_1$ value of its first-shell hydrogens becomes -8 (Table \ref{tab:fractionation_solvation}). This means that H, rather than the heavier isotope D, is favored in the vicinity of the anion. As shown in Fig. \ref{fig:IonDecomp}b, as compared to the monatomic ions, the influence of HPO$_4^{2-}$ extends further into its second solvation layer mainly because the O--H group in the anion is capable of forming hydrogen bonds with these second layer water molecules and disturbs the water-water interactions.

This analysis demonstrates that the QKEs and H/D fractionation ratios provides highly sensitive probes to the ion effects in aqueous solutions. From Table \ref{tab:fractionation_solvation} and Fig. \ref{fig:IonDecomp}, all 3 types of ions have the strongest impact on their first hydration layers. In their vicinity, Na$^+$ perturbs the water structure by not participating in hydrogen bonding interactions, leading to positive $\Delta T_1$ and $\Delta 10^3\ln\alpha_1$. While the anions both result in negative $\Delta T_1$ and $\Delta 10^3\ln\alpha_1$ values, they interact differently with their first hydration layers. Compared to water-water interactions, Cl$^-$ forms weaker hydrogen bonds with water and reduces their QKEs in the rotational and translational DOFs, whereas HPO$_4^{2-}$ has stronger hydrogen bonds with water and decreases their QKE contributions in the vibrational DOFs. As the ion-water distances increase, the solvents are less influenced and $\Delta T_2$ and $\Delta 10^3\ln\alpha_2$ of all the ions become negative, giving an overall decrease in the H/D fractionation ratios as compared to pure water.

\section{Conclusions}
In this work, we have performed AI-PIMD simulations to evaluate the QKEs and isotope fractionation of liquid water and aqueous ionic solutions. By decomposing the total QKE of a water molecule into elements that are associated with the translational, rotational and vibrational DOFs, we are able to demonstrate how the competing quantum effects are modulated by the the condensed phase environment and the ion-water interactions. Our decomposition results could potentially be validated by deep inelastic neutron scattering experiments, which have previously been used to measure the momentum distribution and kinetic energy of atoms in liquids and solids.\cite{Watson1996,Andreani2005,roma+13jpcl,Andreani2019} However, at present these experiments yield information on the kinetic energy anisotropy that is affected by relatively large error bars, so it would not be possible to discriminate the small changes induced by the presence of ions. In addition to the QKEs, We show that the equilibrium isotope fractionation ratios are sensitive probes of the hydrogen bonding environment in liquid water and aqueous solutions. By considering the contributions to the H/D fractionation ratios from solvent molecules in the first solvation shell of the ions, we provide an efficient way to predict the fractionation ratio for a solution of a given concentration, which can be directly compared to the experimental measurements of the equilibrium isotope distributions.

It is well known that NQEs play crucial roles in determining the structure and dynamics of hydrogen bonded systems.\cite{Tuckerman1997,Benoit1998,chen+03prl,Kohen2005,morr-car08prl,Hoefs2009,Wolfsberg2009,Perez2010,Lin2011,liu+13jpcc,ceri+13pnas,Klinman2013,roma+13jpcl,Wang2014a,Pinilla2014,Wilkins2015,Pinotsi2016,Ceriotti2016,Wilkins2017} From analyzing the QKEs and H/D fractionation ratios, both of which arise purely from the quantum mechanical nature of the nuclei, we have uncovered the impact of the three ions on the hydrogen bond network of water. Within the first hydration layer, the cationic Na$^+$ simply breaks the water structures, while the anionic Cl$^-$ and HPO$_4^{2-}$ form hydrogen bonding interactions with the surrounding solvent molecules with different strengths. As the water molecules reside further away from the ions, their properties become more similar to the bulk, and in all cases, addition of ions shifts the balance of competing quantum effects. 
Here we use the BLYP density functional in the AI-PIMD simulations for the purpose of setting up a framework for simulating and analyzing aqueous ionic solutions. While the simulations provide reasonably good predictions of the isotope fractionation ratios as compared to the experimental values, these results could be further improved by using higher tier meta and hybrid exchange correlation functionals that have recently been shown to perform well when used in conjunction with path integral simulations.\cite{Cheng2016,Marsalek2017,RuizPestana2018,Cheng2019} The ability to perform AI-PIMD simulations, which explicitly include both electronic and nuclear quantum effects, allows a detailed understanding of the hydrogen bonding structures and thermodynamic properties of solvated ions, which are of crucial importance to the study of geological and biological systems. 

\begin{acknowledgments}
T.E.M was supported by the National Science Foundation under Grant No. CHE-1652960. T.E.M also acknowledges support from the Camille Dreyfus Teacher-Scholar Awards Program. L.W acknowledges the National Science Foundation through the award CHE-1904800. This work used the Extreme Science and Engineering Discovery Environment (XSEDE), which is supported by National Science Foundation grant number ACI-1548562 (Project TG-CHE170034). M.C acknowledges funding by the European Research Council under the European Union's Horizon 2020 research and innovation programme (grant agreement no. 677013-HBMAP).
\end{acknowledgments}

\end{document}